\documentstyle[11pt,newpasp,twoside,epsf]{article}
\def\edcomment#1{\iffalse\marginpar{\raggedright\sl#1\/}\else\relax\fi}
\reversemarginpar

\begin{document}
\title{Circumbinary disks around T\,Tauri stars: HST/NICMOS
  near-infrared images and polarimetric maps} 
\author{Gaspard Duch\^ene} 
\affil{Observatoire de Grenoble, Universit\'e Joseph
  Fourier, BP 53, 38041 Grenoble Cedex 9, France} 
\author{Joel Silber}
\affil{Department of Physical Sciences, University of Hertfordshire,
  College Lane, Hatfield, Hertfordshire AL10 9AB, UK}
\author{Fran\c{c}ois M\'enard} 
\affil{Canada-France-Hawaii Telescope
    Corporation, PO Box 1597, Kamuela HI 96743, USA} 
\author{Tim Gledhill} 
\affil{Department of Physical Sciences, University of
    Hertfordshire, College Lane, Hatfield, Hertfordshire AL10 9AB, UK}

\begin{abstract}
  We have obtained new near-infrared images of both GG\,Tau and
  UY\,Aur circumbinary disks with the polarimetric modes of NICMOS
  aboard the Hubble Space Telescope. The 1\micron\ intensity map of
  GG\,Tau reveals a more complete elliptical shape than in previous
  ground-based images. Its eastern and western sides are definitely
  asymmetric. Our image strongly supports the ring geometry proposed
  by Guilloteau, Dutrey \& Simon (1999) on the basis of their
  millimetre interferometry images: a geometrically thick and sharply
  edged ring surrounding an empty gap around the binary. Around
  UY\,Aur, we identify structures that are in excellent agreement with
  the optical images of M\'enard et al. (1999), which confirms that
  the inclination of the system to the line-of-sight is about 60
  degrees. We also find tentative new structures closer to the stars.
\end{abstract}

\section{Introduction}

The presence of circumstellar disks around T\,Tauri stars has been
suspected for a long time, but it is only very recently that these
disks were directly detected, using high-angular millimetre imaging.
These images revealed extended gas structures which appeared to be in
Keplerian rotation around the central object. Among the rare
detections so far, two disks were found around binary T\,Tauri stars:
GG\,Tau (Dutrey, Guilloteau \& Simon 1994) and UY\,Aur (Duvert et al.
1998). The separations of the binaries are 0\farcs25 and 0\farcs89
respectively, which correspond to projected physical separations of 35
and 125\,AU at the distance of the Taurus star-forming region
(140\,pc).

In both cases, light scattered off the surface of the disks has been
detected afterwards with adaptive optics imaging at near-infrared
wavelengthes. Roddier et al. (1996) found that the GG\,Tau ring has a
clumpy appearance and that several radial spokes of material extend
from the ring onto the central stars. The ring is brighter in its
northern part, but is detected in all directions. They interpret this
brightness difference as being due to the scattering geometry. The
UY\,Aur case is very different, as Close et al. (1998) only detected
the disk on one side of the binary. Furthermore, they found evidences
that a ``spiral arm'' splits from the main disk and gets closer to the
star. Deconvolution processes were applied in both studies to retrieve
the highest spatial resolution allowed by adaptive optics devices, and
this may lead to some artifacts in the final images.

More recently, the first visible wavelength images of UY\,Aur were
obtained by M\'enard et al. (1999) at 600 and 800\,nm with HST/WFPC2.
The PSF-subtracted images revealed a more complicated structure that
was found by Close et al. (1998): a large ``clump'' appears to be
independent from the disk itself. If true, this implies that the
inclination of the system to the line-of-sight is larger than was
first thought (about 60\deg\ instead of about 40\deg).

To improve our knowledge of these two circumbinary disks, we have
performed new observations at 1\micron\ and 2\micron\ of these systems
with HST/NICMOS. We used the polarimetric modes, and we obtained both
intensity and polarization maps, which do not need to be deconvolved.
The GG\,Tau polarization maps are the first ever obtained of this
system, while Potter et al. (1998) already presented a deconvolved
J-band polarization map of UY\,Aur which revealed a nice
centrosymetric pattern. Polarization maps are powerfull tools to
investigate the dust grain properties and the geometry and structure
of the disks.

In section\,2, we summarize our observations and data processing
steps, and the maps of both systems are presented and commented in
section\,3. Section\,4 describes some implications of our results on
the properties of these disks.

\section{Observations and data processing}

The 1\micron\ and 2\micron\ images were obtained with Camera 1 and
Camera 2 respectively, providing pixel scales of 0\farcs043 and
0\farcs075. Both binaries were observed through the three polarizers at
each wavelength, during three 96 seconds exposures for each filter.
The regular NICMOS data reduction pipeline prooved to be unsatisfying,
and we had to re-reduced all data, with specific care to the so-called
``pedestal effect'', to obtain final images where the sky level is
flat all over the detector.

To allow clear detections of the disks, it is mandatory to remove the
bright stellar point spread funtions (PSFs). We first tried Tinytim
PSFs, but it appeared that their match with the real ones is quite
poor, so we turned to a ``natural'' star, i.e. a bright single star
observed through the same filters. The diffraction spikes subtraction,
though unperfect, is quite good, and the optical ghosts induced by
some polarizers are naturally removed. Some residuals in the core of
the PSFs, however, are still large, and nothing can be securely
detected in the inner 0\farcs5 at 1\micron. At 2\micron, some fringing
can be seen at separations as large as 3\farcs5. No deconvolution
process was applied to our images, which allows an easier
interpretation.

\section{Results}

\subsection{GG\,Tau}

The new 1\micron\ image of the GG\,Tau ring is presented in Fig.\,1.
Its overall geometry is in good agreement with Roddier et al. (1996)'s
images, though with a higher signal-to-noise ratio. However, there are
some noticeable features. First, the ring does not appear clumpy in
our image. This property was likely an artifact introduced by the
deconvolution process applied to the adaptive optics images. Fitting
an ellipse onto the ring, we find a semi-major axis, a position angle
and an inclination in excellent agreement with the millimetre results
of Guilloteau et al. (1999). It is noticeable, however, that this
ellipse is not centered on the center of mass of the binary. Our image
does not allow us to confirm the existence of the spokes of material
discovered by Roddier et al. (1996), because of the large PSF
subtraction residuals inside the ring. Finally, a significant
east-west asymetry in the northern part of the ring is seen in our
intensity map.

\begin{figure}
\plottwo{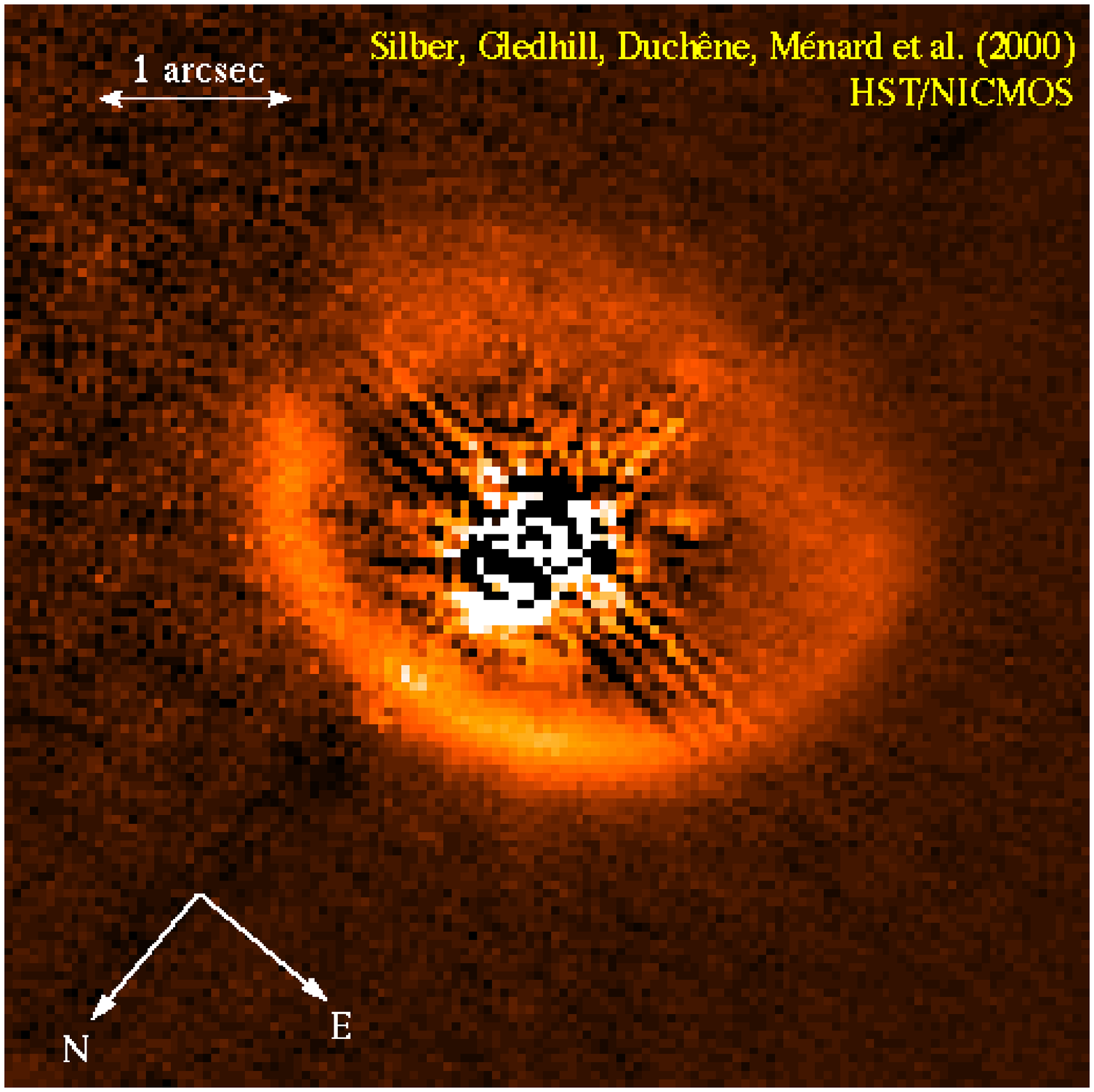}{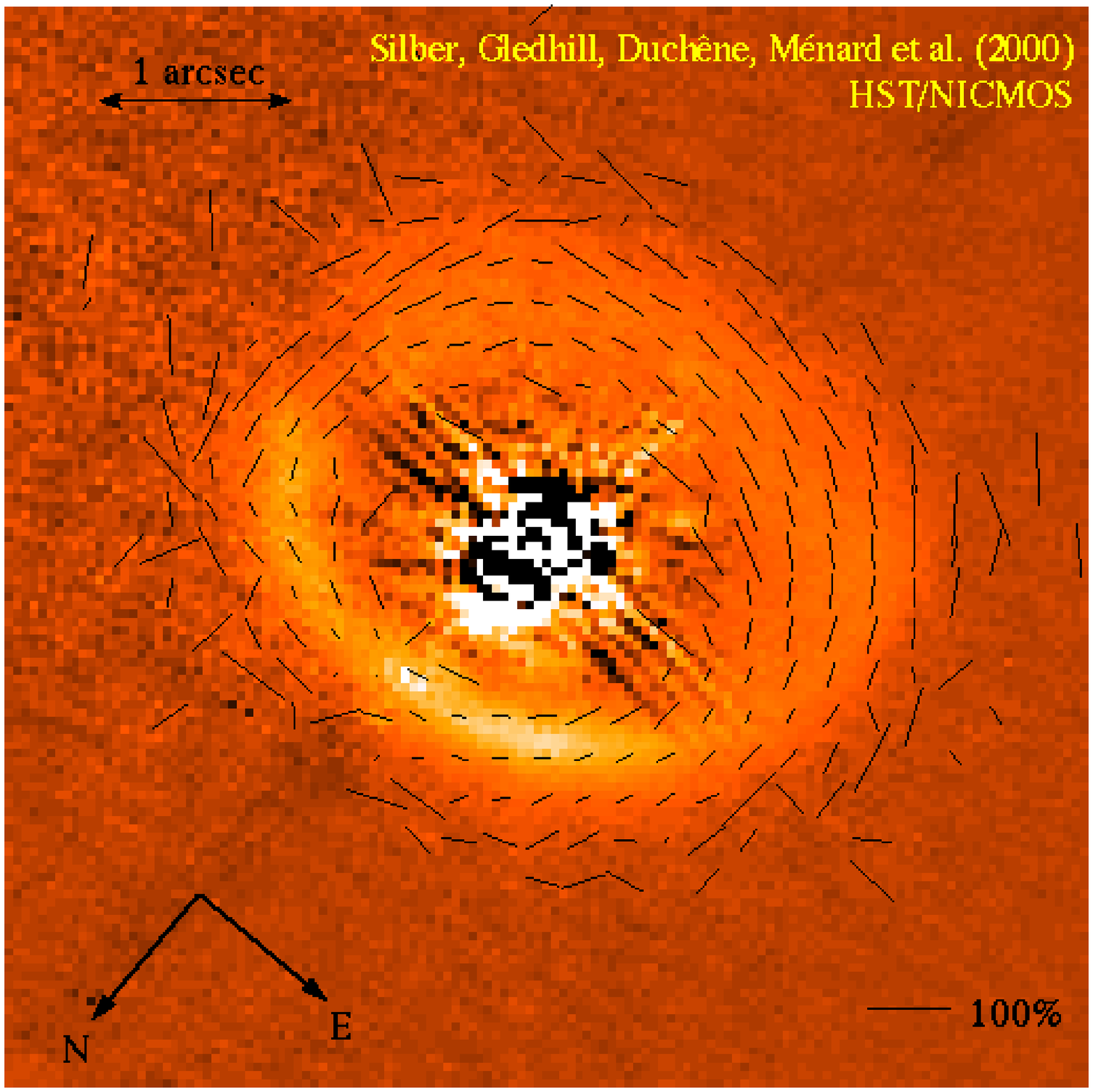}
\caption{PSF-subtracted intensity (left) and polarization (right) maps
  of GG\,Tau at 1\micron.}
\end{figure}

The polarization vectors are strikingly well organized in a
centrosymetric pattern, which is symetric about the semi-minor axis of
the ring. The brightest part of the ring, which is the closest to the
observer, displays a lower polarization level than the faintest side,
typically 20\% as opposed to 50--60\%.

At 2\micron, the disk is too close to the stars, and the large
subtraction residuals prevent us from obtaining a clear image of the
ring. However, we calculated the polarization map at this wavelength
and, though the image is strongly dominated by the unpolarized
stellar fluxes, a centrosymetric pattern is found in the polarization
vectors, with a typical level of 5--10\%, indicating that the
intrinsic polarization level of the light scattered by the ring is
high.

\subsection{UY\,Aur}

The morphology of the UY\,Aur circumbinary disk in our new 1\micron\ 
image is in good agreement with M\'enard et al. (1999)'s optical
image, though the former suffer from a poor signal-to-noise ratio. As
an be seen in Fig.\,2, the disk appears as an unresolved arc to the
Southwest of the binary at both wavelengthes, while a bright clump to
the southeast appears to be unrelated to this structure. Noticeably,
the bright arc seems to widen to the West of the binary in our image.
This can be interpreted as the arc breaking into two separate arcs:
the disk itself remaining at least 2\farcs5 away from the stars, and
an inner arc, getting closer to the stars. The latter would correspond
to the ``spiral arm'' described by Close et al. (1998). A second
feature in our map is an inner arc which is much brighter than the
disk itslef.  It lies about 1\arcsec\ to the Southwest of the
secondary. This may be a PSF artifact, but it can be seen using both a
natural or a Tinytim-built PSF. Furthermore, its coincidence with a
similar arc detected in the WFPC2 images is suggestive (see Fig.\,2).
At 2\micron, the only area which is clearly separated from the PSF
residuals is found to the southeast of the binary. It can be traced up
to 5\arcsec\ away from the stars. The back side of the disk remains
undetected at both wavelengthes, which provides strong constraints on
the dust grain properties. We note however, that the small arc seen to
the Northeast of the primary at 1\micron\ can be seen in all indivual
images.

\begin{figure}
  \plottwo{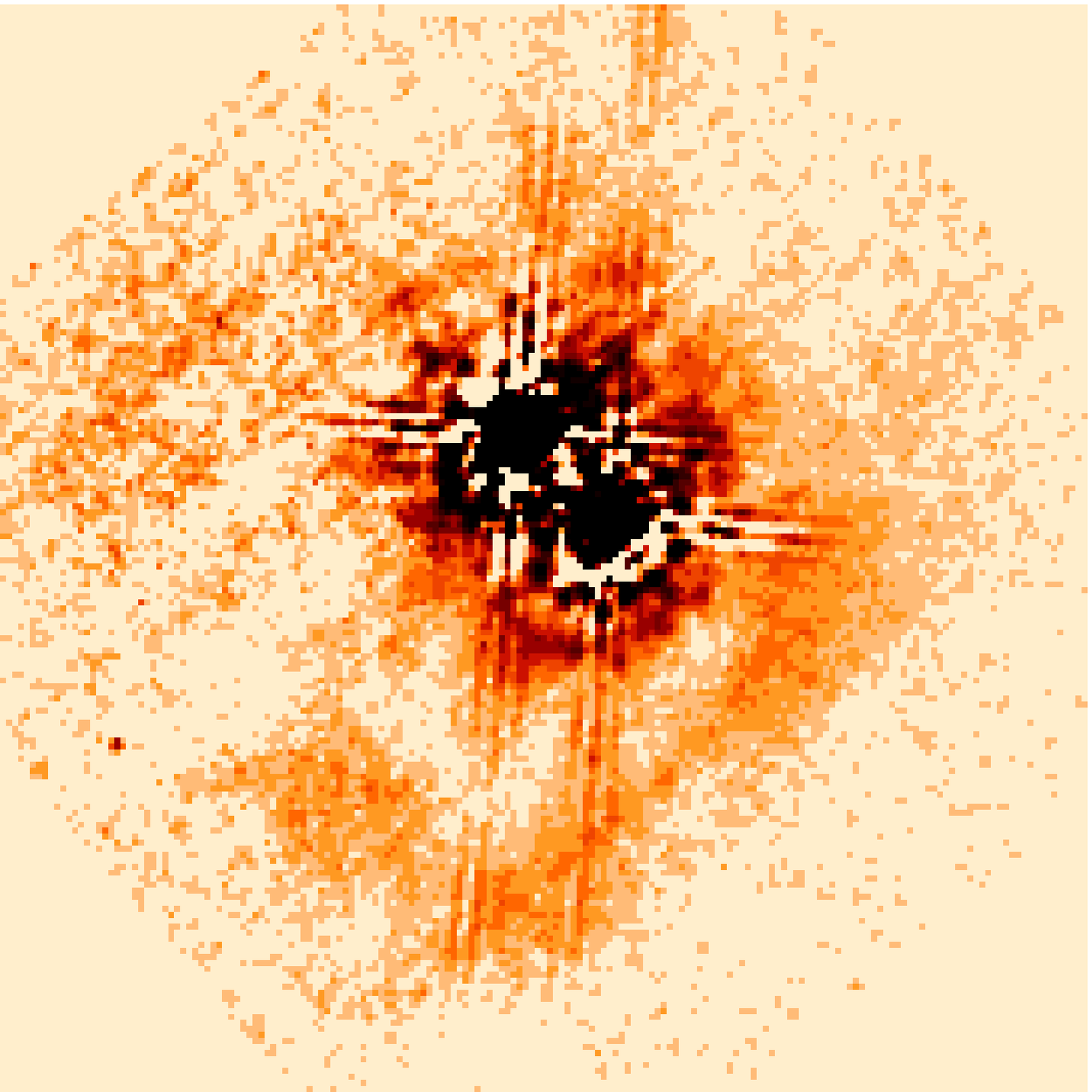}{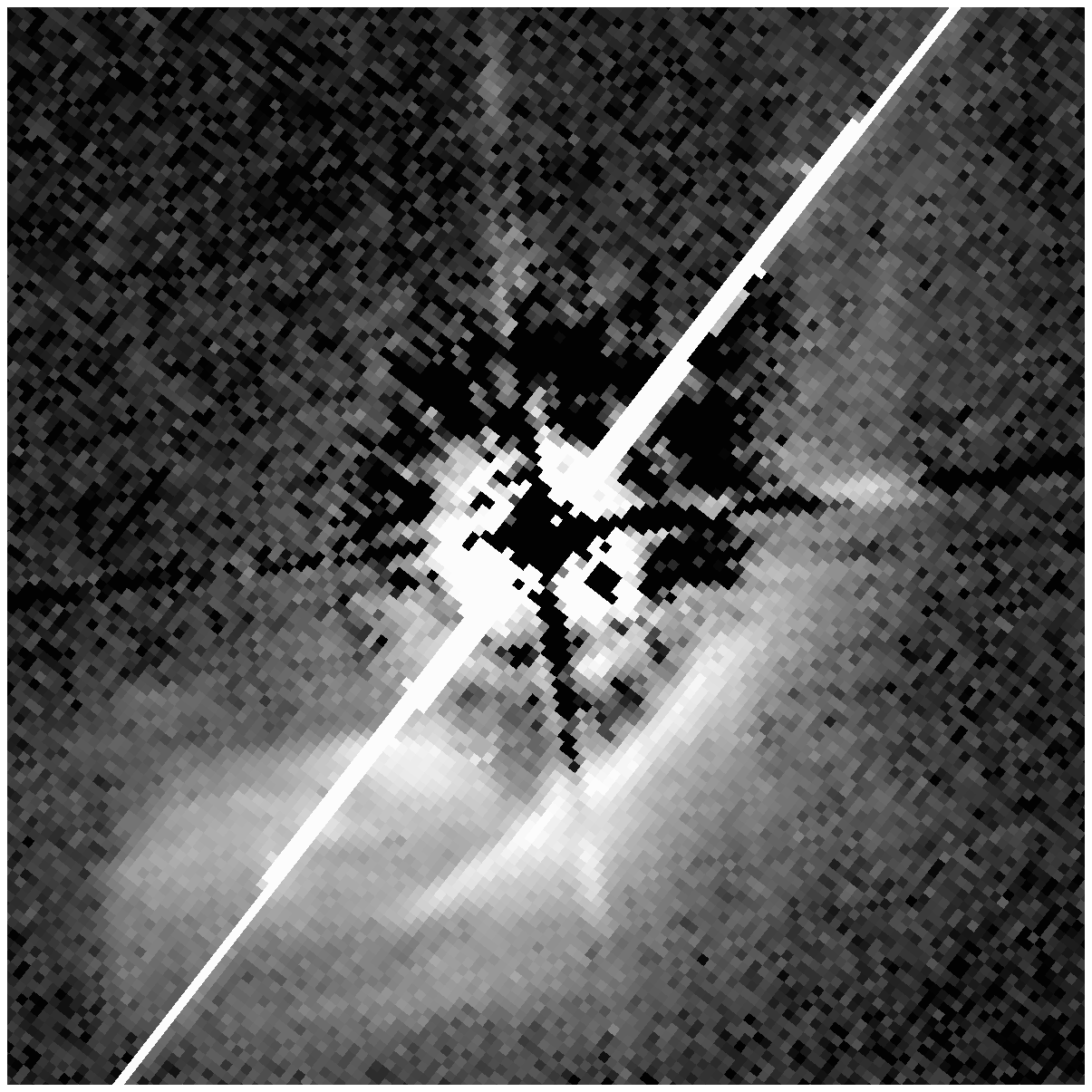}
\caption{Intensity maps of UY\,Aur at 1\micron\ (this work) and at
  600\,nm (from M\'enard et al. 1999). Image sizes are 7\farcs7 and
  12\arcsec\ respectively. North is up, East to the left. The noisy
  area located to the East of the binary in the 1\micron\ image is due
  to a very low response of the detector.}
\end{figure}

\begin{figure}
\plottwo{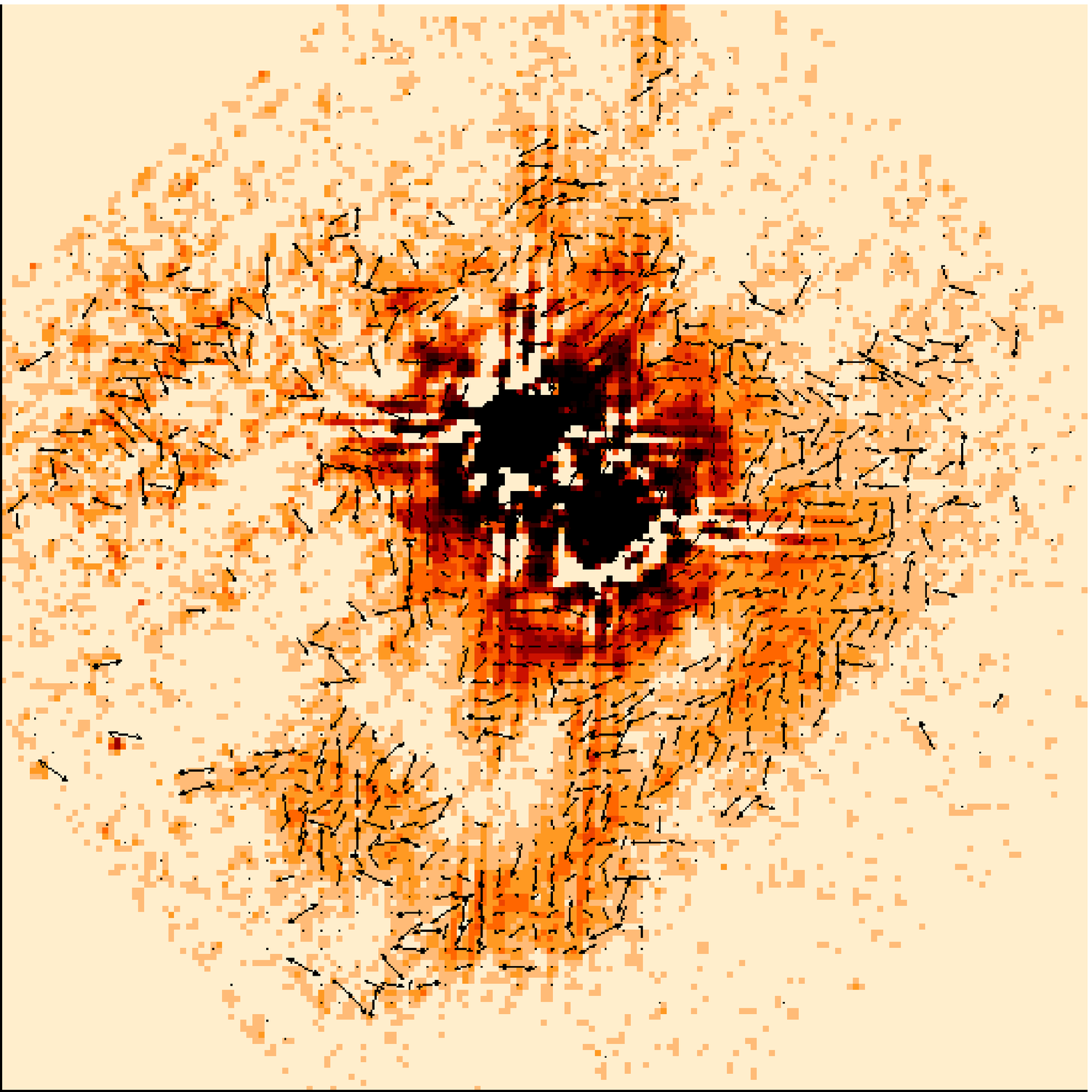}{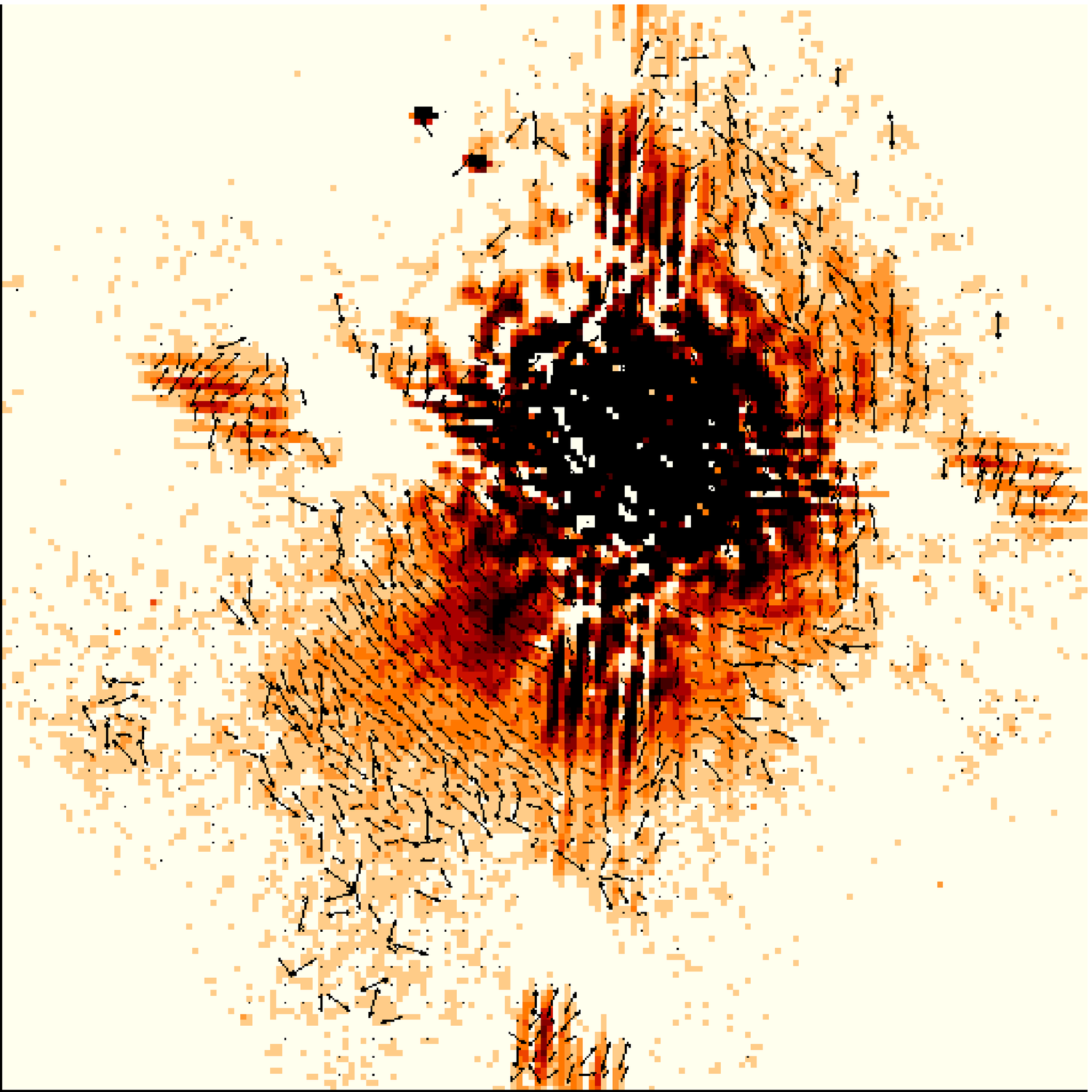}
\caption{Polarization vectors superimposed on the intensity maps of
  UY\,Aur at 1\micron\ (left) and 2\micron\ (right). Image sizes are
  7\farcs7 and 13\arcsec\ respectively. Orientation of both images is
  the same as in Fig.\,2.}
\end{figure}

The polarization pattern at 2\micron\ is well organized in its
southeastern part, with all vectors well aligned, in a fashion
consistent with centrosymetric. The typical polarization level is
about 40\%. At 1\micron, however, the picture is much different:
though the vectors are basically aligned in the southwestern part of
the disk with a typical level of 20\%, again in a more or less
centrosymetrical pattern, they are quite randomly oriented in the
southeastern clump. Whether this is due to our reduction pipeline, to
the low signal-to-noise ratio or reflects the intrinsic pattern is
unclear. It must be noted that Potter et al. (1998) found a very
different behaviour. Their data reduction process, however, included a
deconvolution step, which impact on the polarization is unknwon.

\section{Implications and open questions}

\subsection{GG\,Tau}

As already pointed out by Guilloteau et al. (1999), the shift between
the apparent center of the ring and the center of mass of the binary
is naturally explained by a thick ring geometry. This is related to
the fact that, in the Mie theory, forwards scattering is strongly
favoured. Hence, most of the light scattered towards the observer
comes from the upper part of the disk's inner edge, whose projection
onto the sky is not symetric about the physical center of the ring.
The quantitative model proposed by Guilloteau et al. (1999) is a ring
with a half-thickness of 60\,AU at its inner radius (180\,AU) and a
very sharp edge. The observed location of the ring appears to be in
excellent agreement with the prediction from this model. Roddier et
al. (1996) suggested that the thickness-to-radius ratio was about one
tenth at the inner edge of the ring, which seems imcompatible with our
results.

The origin of the east-west asymetry is unclear. It may be due to the
presence of two illuminating stars, or be related to the slight
asymetry found in the millimetre wavelength image, which itself may
reveal internal structures differences.

\subsection{UY\,Aur}

Fitting ellipses on the large southwest unresolved arc both at
1\micron\ and 600\,nm yields very similar figures and, especially, an
inclination to the line-of-sight of about 60\deg\ in both cases. This
is much larger than the 42\deg\ estimated by Close et al. (1998). The
reason for that discrepancy is that they assumed that the southeast
clump belongs to the disk, which now seems unlikely. Duvert et al.
(1998) pointed out that a larger inclination (60--70\deg) is in better
agreement with the millimetre observations.

If the inner arc close to the secondary star proves to be real, it
will represent a challenge for theoretical studies, as most of them
predict that structures at such a location should be highly unstable
due to the interaction with the binary system.

\subsection{Monte Carlo modelling}

The polarization levels, as well as the front-to-back side flux
ratios, are tightly linked to the disk geometry and to the dust grain
properties. Our next step in this study is to run Monte Carlo
simulations to investigate these properties. For instance, we may
determine whether the dust grain size distribution in the disks is
compatible with that of interstellar grains. In principle, we will
also constrain the amount of ``flaring'' in both disks, as well as
their geometrical height and optical depth.

\end{document}